\documentclass[runningheads]{llncs}
\usepackage[american]{babel}
\usepackage[utf8]{inputenc}
\usepackage[T1]{fontenc}

\usepackage{cite}
\usepackage{amsmath,amssymb,amsfonts}
\usepackage{aas_macros}
\usepackage{algorithm}
\usepackage{algpseudocode}
\usepackage{graphicx}
\usepackage{textcomp}
\usepackage{booktabs}
\usepackage{lipsum}
\usepackage[normalem]{ulem}

\usepackage{url}
\usepackage{hyperref}
 \hypersetup{
 	bookmarksdepth=-2
 } 

\usepackage{multirow,subfig}
\usepackage{makecell}  

\usepackage{color}
\usepackage[dvipsnames]{xcolor}

\begin{document}
\title{Towards a Mini-App for Smoothed Particle Hydrodynamics at Exascale}

\author{Danilo Guerrera\inst{1} \and Rubén M. Cabezón\inst{2}
Jean-Guillaume Piccinali\inst{3} \and Aurélien Cavelan\inst{1} \and Florina M. Ciorba\inst{1} \and David Imbert\inst{4} \and
Lucio Mayer\inst{5} \and Darren Reed\inst{5}}

\institute{Department of Mathematics and Computer Science, University of Basel, Switzerland\\
\email{\{firstname.lastname\}@unibas.ch},
\and
Scientific Computing Center (sciCORE), University of Basel, Switzerland,\\
\email{ruben.cabezon@unibas.ch},
\and
Scientific Computing Support, Swiss National Supercomputing Centre, Lugano, Switzerland,\\
\email{jgp@cscs.ch},
\and
Nextflow Software, Nantes, France,\\
\email{david.imbert@nextflow-software.com},
\and
Center for Theoretical Astrophysics and Cosmology, Institute for Computational Science, University of Z{\"u}rich, Switzerland,\\
\email{\{lmayer,reed\}@physik.uzh.ch}
}

\maketitle

\begin{abstract}
The smoothed particle hydrodynamics (SPH) technique is a purely Lagrangian method, used in numerical simulations of fluids in astrophysics and computational fluid dynamics, among many other fields. 
SPH simulations with detailed physics represent computationally-demanding calculations. 
The parallelization of SPH codes is not trivial due to the absence of a structured grid. 
Additionally, the performance of the SPH codes can be, in general, adversely impacted by several factors, such as multiple time-stepping, long-range interactions, and/or boundary conditions.
This work presents insights into the current performance and functionalities of three SPH codes: 
SPHYNX, ChaNGa, and SPH-flow. 
These codes are the starting point of an interdisciplinary co-design project, SPH-EXA, for the development of an Exascale-ready SPH mini-app. 
To gain such insights, a rotating square patch test was implemented as a common test simulation for the three SPH codes and analyzed on two modern HPC systems. 
Furthermore, to stress the differences with the codes stemming from the astrophysics community (SPHYNX and ChaNGa), an additional test case, the Evrard collapse, has also been carried out. 

This work extrapolates the common basic SPH features in the three codes for the purpose of consolidating them into a pure-SPH, Exascale-ready, optimized, mini-app. 
Moreover, the outcome of this serves as direct feedback to the parent codes, to improve their performance and overall scalability. 

\end{abstract}

\section{Introduction}
\setcounter{footnote}{0}
Understanding how fluids and plasmas behave under complex physical conditions is at the basis of a number of most important questions that researchers try to answer. 
These range from practical solutions to engineering problems, to cosmic structure formation and evolution. 
In that respect, numerical simulations of fluids in astrophysics and computational fluid dynamics (CFD) are among the most computationally-demanding calculations, in terms of sustained floating point operations per second or FLOP/s\footnote{FLOP/s denotes a widely-used and accepted metric for assessing computational performance in scientific computations.}. 
It is expected that these numerical simulations will greatly benefit from the future Exascale computing infrastructures, that will perform $10^{18}$ FLOP/s. 
This type of scenarios pushes the computational astrophysics and CFD fields well into sustained Exascale computing.

The simulation codes used in numerical astrophysics and CFD are numerous and varied. 
Most of such codes rely on a hydrodynamics solver that calculates the evolution of the system to be studied along with all the coupled physics. 
The Smoothed Particle Hydrodynamics (SPH) technique, discussed in Section~\ref{sec:sph}, is one of such hydrodynamics solvers. 
SPH is a purely Lagrangian method, with no subjacent mesh, where the fluid can freely move. 
Additionally, SPH naturally couples with the fastest and most efficient gravity solvers, such as tree-code and fast multiple methods. These aspects make SPH highly convenient for astrophysics and related simulations. In CFD, SPH can be efficiently used to model highly dynamic fluids, without any re-meshing and, coupled to complex boundary conditions.

The parallelization of SPH codes is not straightforward due 
to the lack of a structured particle grid. 
This aspect causes continuously changing interactions between fluid elements (and mechanical structures in CFD), from one time-step to the next.

This work presents insights into the current performance and functionalities of three SPH hydrodynamical codes: SPHYNX~\cite{cabezon2017} that has the focus in Type Ia and Core-Collapse Supernovas, neutron star mergers, and stellar collisions, ChaNGa~\cite{menon2015} that is primarily applied to galaxy and planet formation, and SPH-flow~\cite{oger2016} that focuses on industry-related problems, such as gearbox lubrification or tire splashing. 
These codes have been successfully used in their respective fields for years and are the starting point of SPH-EXA~\cite{SPH-EXA}, an interdisciplinary project involving computer scientists, astrophysicists, and computational fluid dynamics scientists.
The goal of this project is to develop an Exascale-ready SPH mini-app, optimized across multiple levels of parallelism.

An extensive study of the three SPH implementations is performed in this work to gain insights and to expose any limitations and characteristics of the codes, that may represent a major performance degradation today or in the future Exascale era. 
The availability of a smaller sized application (the mini-app), that shares features both on the implementation and the performance sides of a larger and broader code base, is a highly valuable option while moving towards future Exascale platforms. 
The motivating concept behind mini-apps is that all scientific applications are characterized by bottlenecks that affect their overall performance. 
Thus, it is of high significance to provide a mini-app that fully represents the parent applications by synthesizing their characteristics, but at the same time is easier to handle and whose usage does not require the user to be an expert of the code(s). 
Moreover, mini-apps are a good compromise (in terms of size) for being used in simulators.
This facilitates the study of their performance on various processor architectures, network architectures, and scalability studies beyond the typical practice.

This work presents in Section~\ref{sec:experiments} first results obtained via performance profiling of the parent codes.
The analysis allows to pinpoint which parts of the parent codes are responsible for particular execution behaviors. 
In particular, it allowed to identify the regions where most of the time is spent in, or the impact of communication on the total execution time.

The present analysis and testing of the three parent SPH codes forms the basis for the (future) design of efficient parallelization methods and fault-tolerance mechanisms to implement in the SPH-EXA mini-app.

The remainder of this work is organized as follows. 
Existing methods, efforts, and best practices in developing mini-applications are reviewed in Section~\ref{sec:related-work}. 
The smoothed particle hydrodynamics method is described in Section~\ref{sec:sph}. 
The approach taken in this work towards a co-designed SPH-EXA mini-app is detailed in Section~\ref{sec:towards-mini-app}.
Section~\ref{sec:experiments} includes experiments for two common test simulation together with a performance analysis of the three parent SPH codes. 
The work concludes and outlines the next steps in Section~\ref{sec:conclusion}. 


\section{Related Work}
\label{sec:related-work}
Mini-apps or proxy-apps have received great attention in recent years, with several projects being developed or under development. 
In the high performance computing (HPC) community, the Mantevo Suite~\cite{Heroux2009}, at Sandia National Laboratory, represents one of the first large mini-app set. 
The Mantevo Suite includes mini-apps designed to represent the performance of finite-element codes, molecular dynamics, and contact detection, to name a few. 

In the oceanographic community, CGPOP~\cite{CGPOP} is another example of a mini-app developed to ensure performance portability and testing of new programming models, and implements a conjugate gradient solver that represents the bottleneck of the full Parallel Ocean Program application. 

MCMini~\cite{MCMini} is a co-design application for Exascale research, developed at Los Alamos National Laboratory, that implements Monte Carlo neutron transport in OpenCL and targets accelerators and coprocessor technologies. 

The CESAR Proxy-apps~\cite{CESAR} represent a collection of mini-apps belonging to three main classes: thermal hydraulics (for fluid codes), neutronics (for neutronics codes), and coupling and data analytics (for data-intensive tasks).

One of the motivation behind the European ESCAPE project~\cite{ESCAPE} is to define and encapsulate the fundamental building blocks (`Weather 
\& Climate Dwarfs') that underlie weather and climate services. 
This serves as a prerequisite for any subsequent co-design, optimization, and adaptation efforts. 
One of the ESCAPE outcomes is Atlas~\cite{Atlas}, a library for numerical weather prediction and climate modeling, with the primary goals to exploit the emerging hardware architectures forecasted to be available in the next few decades.
Interoperability across the variegated solutions that the hardware landscape offers is a key factor for an efficient software and hardware 
co-design~\cite{Schulthess2015}, thus of great importance when targeting Exascale systems.

Similar to these works, the creation of a mini-app directly from existing codes rather than the development of a code that mimics a class of algorithms has been recently discussed~\cite{Messer2015}. 
A scheme to follow was proposed therein that must be adapted according to the specific field the parent code originates in.
To maximize the impact of a mini-app on the scientific community, it is important to keep the build and run system easy enough, to not discourage potential users. 
The building should be kept as simple as a Makefile and the preparation of the run to a handful of command line arguments: 
``\textit{if more than this level of complexity seems to be required, it is possible that the resulting MiniApp itself is too complex to be human-parseable, reducing its usefulness.}" \cite{Messer2015}.
The present work introduces the interdisciplinary co-design of an SPH-EXA mini-app with three parent SPH codes originating in the astrophysics academic community and the industrial CFD community.
This represents a category not discussed in~\cite{Messer2015}. 

Skeleton applications, the name used to refer to reduced versions of applications that produce the same network traffic of the full ones, are of interest to model the performance of networks through simulation. An auto-skeletonization approach has recently been proposed to auto-skeletonize a full application via compiler pragmas: an interesting idea to produce flexible skeletons, that serve as a tool to study balanced Exascale interconnect designs~\cite{Wilke:2018}. Such a skeletonization approach may be of interest in annotating a full application to obtain a mini-app that reflects exactly the corresponding production code.

\section{Smoothed Particle Hydrodynamics}\label{sec:sph}

Smoothed particle hydrodynamics (SPH) is a fully Lagrangian meshless method to perform hydrodynamical simulations. 
Initially devised in the late '70s~\cite{lucy1977,gingold1977}, this technique underwent sustained development ~\cite{cabezon2008,garciasenz2012,price2012,rosswog2015,letouze2008,oger2016hpc} and is, at the time of writing, common in many fields, including computational fluid dynamics, plasma physics, solid mechanics, and astrophysics.
Its inherent good conservation properties, and its adaptability to distorted geometries, make SPH a common choice to simulate highly-dynamic three-dimensional (3D) scenarios.

The SPH technique discretizes a fluid in a series of interpolation points (SPH particles) whose distribution follows the mass density of the fluid, and their evolution relies on a weighted interpolation over close neighboring particles. 
This has several implications:
\begin{itemize}
	\item Physical properties are smoothed within the range of the interpolation, named \textit{smoothing length} ($h$). 
 	This smoothing length determines the local spatial resolution.
	\item The identity of the neighboring particles change as the simulated system evolves. 
	There is no (computationally-efficient) method to predict which particles will be neighbors over time. Thus, the neighbors are identified in every time-step.	
	\item To avoid the $\mathcal{O}(N^2)$ calculation incurred by identifying neighbors, a tree structure is typically employed, with the Barnes-Hut tree~\cite{barnes1986} being one of the most common. 
\end{itemize}

Depending on the scenario they simulate or the research field in which they are used, SPH codes can greatly vary in terms of implementation and physical processes that they include. 
Nevertheless, many SPH codes apply the same general underlying workflow, such as the one listed in Algorithm~\ref{fig.algo.sph}.

\begin{algorithm}

\caption{SPH General Computational Workflow}
\label{fig.algo.sph}
\begin{algorithmic}
 \State Initialization
 \While{Target simulated time is not reached}
 	\State 1. Build tree
 	\State 2. Find neighbors and smoothing length
 	\State 3. Execute SPH- and physics-related kernels
 	\State 4. (Optional) Compute self-gravity
 	\State 5. Compute new time-step
 	\State 6. Update velocity and position
 \EndWhile
\end{algorithmic}

\end{algorithm}

As a general rule, once the particle positions and masses are known, a tree is built in step~1 in Algorithm~\ref{fig.algo.sph} and is walked in step~2 to identify the neighbors\footnote{The simulation will try to reach a given target number of neighbors and this influences the value of the resulting smoothing length.} that will be used for the remainder of the calculations in step~3. 
These include the evaluation of the particles' density, acceleration, and rate of change of internal energy. 
Also, in general, the physical modules relevant to the studied scenario are usually located in this step or close to it, as represented by step~4, where self-gravity is calculated. 
Next, a new physically relevant and numerically stable time-step is found (step~5) and the particles are updated (step~6).

As the simulated scenarios grow in complexity and require increased resolution and accuracy, the overall number of particles in SPH simulations increases, as well as the number of neighbors. 
At the time of writing, 3D SPH simulations require between $10^5 - 10^{12}$~particles, with $\sim 10^2$ neighbors per particle.

\section{Towards a Co-designed SPH-EXA Mini-App}
\label{sec:towards-mini-app}

The long-term goal of SPH-EXA~\cite{SPH-EXA} is to provide a parallel, optimized, state-of-the-art implementation of basic SPH operands with classical test cases used by the SPH user community. 
Optimization is critical to achieve the scalability needed to exploit Exascale computers. 
This can be implemented at different levels: employing state-of-the art dynamic load balancing algorithms, fault-tolerance techniques, programming languages, tools, and libraries.

In reaching this goal, interdisciplinary co-design and co-development is the recommended approach for developing a mini-app, to leverage the involvement of the developers of the parent codes in the mini-app design and implementation process~\cite{Heroux2009}. 
Regarding the co-design of the present \mbox{SPH-EXA} mini-app both, computer scientists and computational scientists (that developed the three parents codes) are part of the team.

Based on the co-design principle, we individuated the common and best features of the three parents codes.
First, Table~\ref{table:differences_scientific} summarizes the main differences and similarities of the codes with a focus on the physics therein. 
While not all existing techniques and algorithms need to be implemented in the mini-app, some of them, such as the SPH interpolation kernels, can be implemented as separate interchangeable modules. 
In that sense, Table~\ref{table:mini-app-scientific} shows the domain science techniques and algorithms that the mini-app will feature.

Then, Table~\ref{table:differences_cs} reveals the computer science-related similarities and differences of the codes, i.e., algorithms, techniques, and other implementation-specific choices. 
Each code has a different history, has been used for different purposes, and, therefore, uses different approaches in its simulations. 
For example, all applications use standard checkpoint/restart mechanisms to enable fault-tolerance when executing at scale. 
Yet, they all use different domain-decomposition methods and scheduling techniques.
Table~\ref{table:mini-app-cs} presents the computer science-related features to be implemented in the \mbox{SPH-EXA} mini-app. 
It is important to note that such features can dramatically affect the scalability of the application, as shown in Section~\ref{sec:experiments}.
Therefore, the mini-app design goal is to select state-of-the start techniques that can be applied to SPH to achieve Exascale-level scalability.

\begin{table*}[t]
\centering
\caption{Differences and similarities between SPH-flow, SPHYNX, and ChaNGa} 
\label{table:differences_scientific}
\resizebox{\textwidth}{!}
{
\begin{tabular}{@{}lllllllll@{}}
\toprule
 \textbf{SPH} & \textbf{Code}       & \multirow{2}{*}{\textbf{Kernel}}  & \textbf{Gradients}     & \textbf{Volume} & \textbf{Mass of}  & \textbf{Time-} & \textbf{Neighbour} & \multirow{2}{*}{\textbf{Self-Gravity}} \\
\textbf{Code } & \textbf{Version}       &  & \textbf{Calculation}     & \textbf{Elements} & \textbf{Particles}  & \textbf{Stepping} & \textbf{Discovery} &                                  \\ \midrule
SPHYNX   & 1.3.1 & Sinc     & IAD                       & Generalized     & Equal or Variable             & Global        & Tree Walk            & Multipoles (4-pole) \\
ChaNGa   & 3.3 & \makecell[l]{Wendland,\\ M4 spline} & Kernel derivatives & Standard               & Equal or Variable           & Individual    & Tree Walk            & Multipoles (16-pole)\\
SPH-flow & 17.6 & Wendland & Kernel derivatives  & Standard               & Equal or Adaptive & Global        & Tree Walk            & No                                       \\
\bottomrule
\end{tabular}
}
\end{table*}

\begin{table*}[t]
\centering
\caption{Outlook on the scientific characteristics of the future SPH-EXA mini-app}
\label{table:mini-app-scientific}
\resizebox{\textwidth}{!}
{
\begin{tabular}{@{}llllllll@{}}
\toprule
\textbf{SPH-EXA}        & \makecell[l]{\textbf{Kernel}}   &\makecell[l]{\textbf{Gradients} \\ \textbf{Calculation}}   	& \makecell[l]{\textbf{Volume} \\ \textbf{Elements}} 	& \makecell[l]{\textbf{Mass of} \\ \textbf{Particles}}  	& \makecell[l]{\textbf{Time-} \\ \textbf{Stepping}} & \makecell[l]{\textbf{Neighbour} \\ \textbf{Discovery}} & \makecell[l]{\textbf{Self-Gravity}} \\ \midrule
\makecell[l]{mini-app}   & \makecell[l]{Sinc, M4 spline,\\Wendland}     & \makecell[l]{IAD, \\ Kernel derivatives}                        & \makecell[l]{Generalized, \\ Standard}       & \makecell[l]{Equal, Variable, \\ and Adaptive}               & \makecell[l]{Global, \\ Individual}        & Tree Walk            & Multipoles (16-pole) \\ \bottomrule
\end{tabular}
}

\end{table*}

\begin{table*}[t!]
\centering
\caption{Different and similar computer science-related aspects between SPH-flow, SPHYNX and ChaNGa}
\label{table:differences_cs}
\resizebox{\textwidth}{!}{
\begin{tabular}{@{}lllcrllr@{}}
\toprule
 \textbf{SPH}        & \textbf{Domain}  		& \textbf{Load}     		& \textbf{Checkpoint-} & \multirow{2}{*}{\textbf{Precision}}  & \multirow{2}{*}{\textbf{Language}} & \multirow{2}{*}{\textbf{Parallelization}} & \multirow{2}{*}{\textbf{\#LOC}}  \\
\textbf{Code }        & \textbf{Decomposition} & \textbf{Balancing}     & \textbf{Restart}				 & 				  & 				& &            \\ \midrule
SPHYNX   & Straightforward     		& None (static)                      	   & Yes 		    	& 64-bit     & Fortran 90,  & MPI+OpenMP      & 25,000             \\
ChaNGa   & Space Filling Curve 		& Dynamic			&  Yes          		 & 64-bit    & C++  & MPI+OpenMP+CUDA & 110,000 \\
SPH-flow & Orthogonal Recursive Bisection & Local-Inner-Outer  		& Yes               	& 64-bit  & \makecell[l]{Fortran 90} & MPI  & 37,000 \\ \bottomrule
\end{tabular}
} 
\end{table*}

\begin{table*}[t]
\centering
\caption{Outlook on the computer science features of the future SPH-EXA mini-app}
\label{table:mini-app-cs}
\resizebox{\textwidth}{!}{%
\begin{tabular}{@{}lllllllc@{}}
\toprule
\textbf{SPH-EXA}       & \makecell[l]{\textbf{Domain} \\ \textbf{Decomposition}} & \textbf{Parallelization} & \makecell[l]{\textbf{Load} \\ \textbf{Balancing}}     & \makecell[l]{\textbf{Checkpoint-} \\ \textbf{Restart}} & \makecell[l]{\textbf{Error} \\ \textbf{Detection}} & \textbf{Precision}  & \textbf{Language} \\ \midrule
\makecell[l]{mini-app}   & \makecell[l]{Orthogonal Recursive Bisection, \\ Space Filling Curves} &  \makecell[l]{\makecell[c]{X+Y+Z}\\ X=\{MPI\} \\Y=\{OpenMP, HPX\} \\ Z=\{OpenACC, CUDA\}}     & \makecell[l]{DLB with \\ self-scheduling \\ per X, Y, Z level}                        & \makecell[l]{Optimal interval \\ Multilevel}     &  \makecell[l]{Silent data cor-\\ruption detectors} & \makecell[l]{64-bit}               & \makecell[l]{C++}\\ \bottomrule
\end{tabular}
} 
\end{table*}

The mini-app will provide a reference implementation in MPI+X. MPI is the de facto standard for HPC applications, due to the lack of a valid alternative for the inter-node communication.
The MPI+OpenMP programming model does not fully exploit the heterogeneous parallemlism in the newest architectures. While OpenMP 4.5~\cite{OpenMP} offers support for accelerators, in the meanwhile other languages directly targeting accelerators have being proposed and accepted by the community, such as OpenACC (a directive-based programming model targeting a CPU+accelerator system, similar to OpenMP), CUDA (an explicit programming model for GPU accelerators) and  OpenCL. In programming models, research has been focusing on the efficient use of intra-node parallelism, able to properly exploit the underlying communication system through a fine grain task-based approach, ranging from libraries (Intel TBB~\cite{TBB}) to language extensions (Intel Cilk Plus~\cite{CilkPlus} or OpenMP), to experimental programming languages with focus on productivity (Chapel~\cite{Chapel}). Kokkos~\cite{Kokkos} offers a programming model, in C++, to write portable applications for complex manycore architectures, aiming for performance portability. HPX~\cite{HPX} is a task-based asynchronous programming model that offers a solution for homogeneous execution of remote and local operations. It is planned to port the mini-app to at least one of the paradigms described, between the ones described previously to explore their efficiency and potential on Exascale ready machines.

To enable the scalable execution of SPH codes, massive software parallelism needs to be exposed and expressed during parallelization. 
This will require (hierarchical) dynamic load balancing (DLB) techniques to exploit the massive hardware parallelism. 
The SPH-EXA mini-app will employ algorithms, techniques, and tools that address the load imbalance factors arising from the (problem and algorithmic) characteristic of the three SPH codes (multi-time-stepping) as well as from the software environments (processor speed variations, resource sharing). 
This will minimize the load imbalance between synchronous parts of the code (e.g. gravity calculation) by dynamically distributing the load to the processors, using state-of-the-art load balancing methods~\cite{FRAC:1996,Banicescu2003-scalability,ciorba:2006}.

Furthermore, fault, errors and failures have become the norm rather than the exception in large-scale systems~\cite{Snir_FailuresExascale,IESP-toward,JSFI14}. Providing adequate fault-tolerance mechanisms has become mandatory. While checkpointing, rollback and recovery~\cite{CL85,uncoordinated} is the de-facto general-purpose recovery technique to tolerate failures during the execution, optimal multilevel checkpointing can leverage the different storage mediums available to greatly enhance performance~\cite{Di16_Twolevel,benoit16_ipdps}.
Additional mechanisms to handle silent errors, or silent data corruptions must also be considered~\cite{BGomez:2016,Sridharan:2015:MEM:2694344.2694348}.

While many older HPC codes were originally written in Fortran, C++ has become a common choice for developing new projects, including mini-apps such as LULESH~\cite{osti_1090032} and the proposed SPH-EXA mini-app.
The complexity and diversity of the hardware architectures supported by C++ steadily increases, and it also allows implementing the proper abstractions to make applications sustainable in the future.
In particular, it allows the development of type-safe, flexible and portable functionalities, with no runtime overhead.

Reproducibility is at the core of the scientific method, its cornerstone being the ability to independently reproduce and reuse experimental results to prove and build upon them. 
While the concept of reproducibility has always been part of the science, in certain computational sciences it has long been neglected. 
A framework was proposed to support reproducible research~\cite{Guerrera2018} and will be analyzed, evaluated, and adopted as in the design of the SPH-EXA mini-app.
Two test cases were implemented that will also serve as validation and acceptance proofs for the SPH-EXA mini-app (see Section~\ref{sec:experiments}).

\section{Analysis and Testing of the SPH Codes}
\label{sec:experiments}
The complexity of the scenarios simulated in CFD and Astrophysics usually forbids the possibility to perform simulations with continuously increased resolution and different codes, so that a convergence to zero differences on the results can be found. 
Often, it is neither possible nor reasonable to obtain sufficient computational resources to perform simulations that are ``converged'' throughout the computational domain in a mathematical sense. 
It is much more important to limit the deviations in under-resolved regimes by enforcing fundamental conservation laws. 
Additionally, the emergence of stochastic processes, like turbulence, renders the pursuit for mathematical convergence impossible, yet still constrained by conservation laws. 
As a consequence, overall physics properties of the simulated scenarios remain robust, even if slightly different results are obtained when using different codes to ``solve'' the same set of equations. 
Therefore, comparing results of different hydrodynamical codes to the same initial conditions has been proved to be highly beneficial to gain understanding in complex scenarios, in the behavior of the codes, and to discover strengths and weaknesses of those. 
These comparisons are not uncommon in CFD and Astrophysics~\cite{liebendoerfer2005, agertz2007, tasker2008, cabezon2018}.
The SPH codes comparison presented herein is in this spirit, yet with a focus on performance-related aspects from computer science.
			
\subsection{Simulation Tests}

\begin{table*}
    \centering
\caption{Test simulations and their characteristics}
\label{table:differences_simulations}
\resizebox{\textwidth}{!}{
\begin{tabular}{@{}llcccr@{}}
\toprule
 \multirow{2}{*}{\textbf{Test Simulation}}        	  & \multirow{2}{*}{\textbf{Description}}	&  \multirow{2}{*}{\textbf{Domain Size}}    		& \textbf{Simulation} &  \multirow{2}{*}{\textbf{SPH Code}}  & \textbf{Test} \\
								        & 								 & 									& \textbf{Length}	& 							 & \textbf{Platform} \\ \midrule
 \multirow{2}{*}{\makecell[l]{Rotating Square Patch~\cite{colagrossi2005}}}   & Rotation of a free-surface 	&  \multirow{2}{*}{3D, $10^{6}$ particles}  &  \multirow{2}{*}{20 time-steps}    	& \multirow{2}{*}{\makecell{SPHYNX, ChaNGa \\ SPH-flow}}& \multirow{5}{*}{\makecell[r]{Piz Daint \\ MareNostrum 4}}\\
							      & square fluid patch	&                       	   				& 										&  					&\\
 & & & & & \\
 \multirow{2}{*}{\makecell[l]{Evrard Collapse~\cite{evrard1988}}}             & Adiabatic collapse of an initially cold				&  \multirow{2}{*}{3D, $10^{6}$ particles}  &  \multirow{2}{*}{20 time-steps} 	    	& \multirow{2}{*}{SPHYNX, ChaNGa} & \\
							      & and static gas sphere (w/ self-gravity)				&                       	   				& 								& \\
							      \bottomrule
\end{tabular}
} 
\vspace{-0.5\baselineskip}
\end{table*}

\subsubsection{Rotating square patch}
This test was first proposed by~\cite{colagrossi2005} as a demanding scenario for SPH simulations. 
The presence of negative pressures stimulates the emergence of unphysical tensile instabilities that destroy the system. 
Nevertheless, these can be suppressed either using a tensile stability control or increasing the order of the scheme~\cite{oger2016}. 
As a consequence, this is a commonly used test in CFD to verify hydrodynamical codes, and it is employed it in this work as a common test for the three codes.

The setup here is similar to that of \cite{colagrossi2005}, but in 3D. 
The original test was devised in 2D, but the SPH codes used in this work normally operate in 3D. 
To use a test that better represents the regular operability of the target codes, the square patch was set to $[100 \times 100]$ particles in 2D and this layer was copied 100 times in the direction of the Z-axis. 
This results in a cube of $10^6$ particles that, when applying periodic boundary conditions in the Z direction, is equivalent to solving the original 2D test 100 times, while conserving the 3D formulation of the codes. 
The initial conditions are the same for all layers, hence they depend only on the X and Y coordinates. 
The initial velocity field is given such that the square rotates rigidly:
\begin{equation}
v_x(x,y)=\omega y;\qquad v_y(x,y)=-\omega x,
\label{eq:square_v0}
\end{equation}
\noindent
where $v_x$ and $v_y$ are the X and Y coordinates of the velocity, and $\omega = 5$~rad/s is the angular velocity. 
The initial pressure profile consistent with that velocity distribution can be calculated from an incompressible Poisson equation and expressed as a rapidly converging series:
%
\begin{multline}{\notag}
P_0=\rho\sum_{m=0}^\infty\sum_{n=0}^\infty\frac{-32\omega^2}{mn\pi^2\left[\left(\frac{m\pi}{L}\right)^2 + \left(\frac{n\pi}{L}\right)^2 \right]} \times  \\
\sin\left(\frac{m\pi x}{L}\right)\sin\left(\frac{n\pi x}{L}\right),
\end{multline}
\label{eq:square_p0}
%
\noindent
where $\rho$ is the density and $L$ is the side length of the square.

\subsubsection{Evrard collapse}
The Evrard collapse~\cite{evrard1988} is a test commonly used to verify astrophysical codes. 
It consists of an initially static isothermal spherical cloud of gas (mimicking a star) that undergoes an accelerated gravitational collapse, until the rapid rise of temperature and pressure at its core produces a shock wave that expands from the center of the star to its outer layers. 
The Evrard collapse involves ingredients that are capital for astrophysical simulations, namely shock waves and self-gravity.

The initial conditions were obtained by following the configuration of~\cite{cabezon2017}. 
The initial density radial profile was given by:
\begin{equation}
\rho(r)=
\begin{cases}
 M/(2\pi R^2 r) &\text{for $r \le R$,}\\
 0 &\text{otherwise,}
\end{cases}
\label{evrardprofile}
\end{equation}
\noindent
where $R=1$ and $M=1$ are the initial total radius and mass of the star, respectively. 
The initial internal energy was set to $u_0=0.05$ per unit of mass and an ideal equation of state with $\gamma=5/3$ was used. 
With this configuration the gravitational energy is much larger than the internal energy and the system collapses naturally.

As this test needs the evaluation of self-gravity, it was only performed by the astrophysical SPH codes SPHYNX and CHaNGa. 
We used $10^6$ particles in all simulations of this test.

\subsection{Experimental Results}

\subsubsection{System overview - Piz Daint}
The experiments were performed on the hybrid partition of the Piz Daint\footnote{\url{https://www.cscs.ch/computers/piz-daint/}} supercomputer with Cray MPICH 7.6.0, OpenMP 4.0 and Charm++ 6.8.2
At the time of writing, this supercomputer consisted of a multi-core partition of 1,813 Cray XC40 nodes with two Intel Xeon E5-2695 v4 (codename Broadwell) processors, which were not used in this study, as well as the hybrid partition of 5,320 Cray XC50 nodes. 
These hybrid nodes are equipped with an Intel E5-2690 v3 CPU (codename Haswell) and a PCIe version of the NVIDIA Tesla P100 GPU (codename Pascal) with 16 GB second generation high bandwidth memory (HBM2).
The nodes of both partitions are interconnected in one fabric based on Aries technology in a Dragonfly topology\footnote{\url{http://www.cray.com/sites/default/files/resources/CrayXCNetwork.pdf}}.

\subsubsection{System overview - MareNostrum 4}
The experiments were also performed on the MareNostrum\footnote{\url{https://www.bsc.es/marenostrum/marenostrum}} supercomputer with Intel MPI 2017.3 and OpenMP 4.5.
At the time of writing, this supercomputer consisted of a partition of 3,456 Lenovo nodes with two Intel Xeon Platinum 8160 (codename Skylake) processors. 
The nodes are interconnected with 100Gb Intel Omni-Path in a Full-Fat Tree topology.

\subsubsection{Analysis of strong scalability}

Generating initial conditions for different numbers of particles is a \mbox{non-trivial} process.
Therefore, this work employs a set of strong-scaling experiments to assess the performance at scale with fixed number of particles for each test.

\begin{figure*}[h!]%
\centering
\subfloat[]{
  \includegraphics[width=0.4\textwidth]{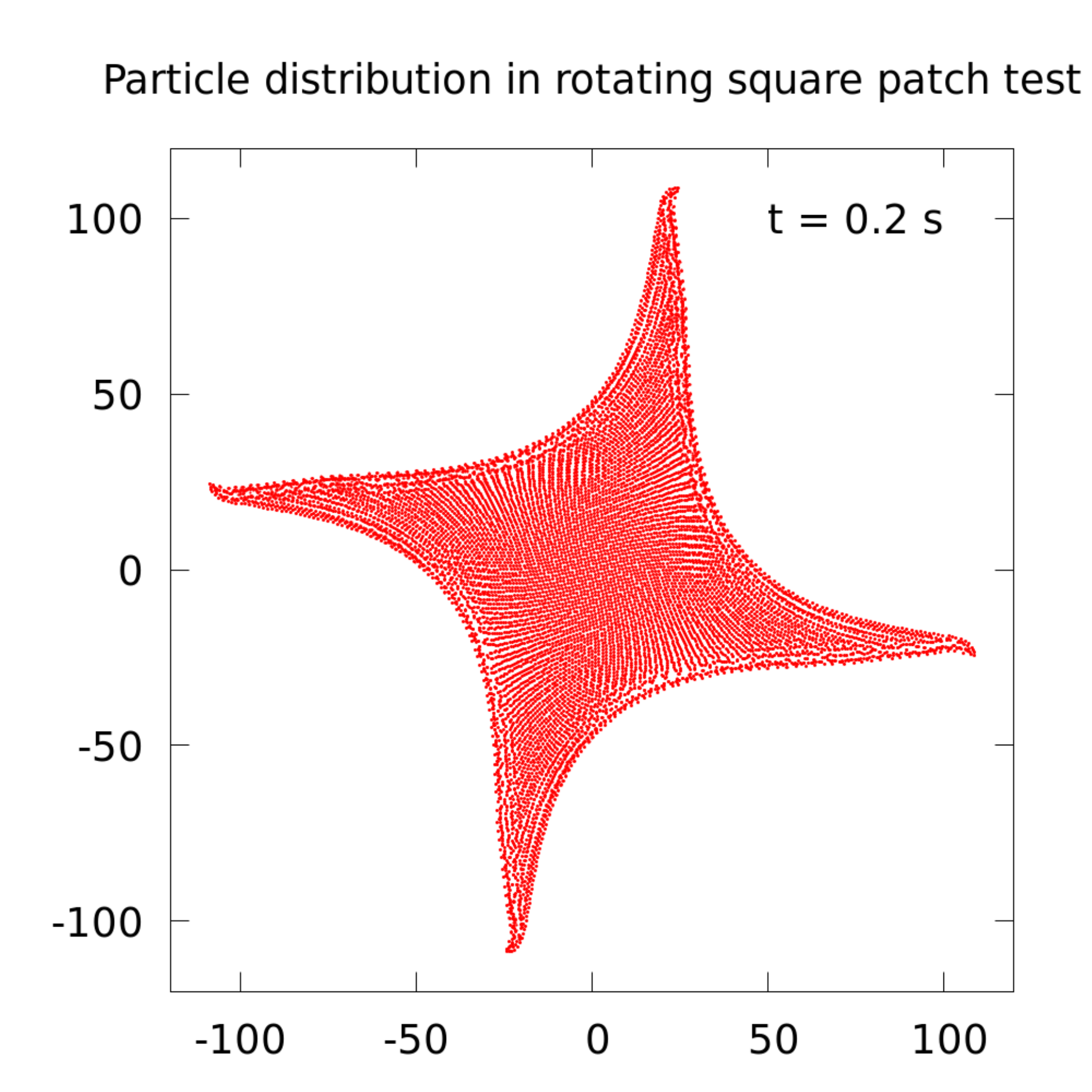} 
  \label{tab:sphynx_pop}
}%
\hspace{0.5cm}
\subfloat[]{
  \includegraphics[width=0.5\textwidth]{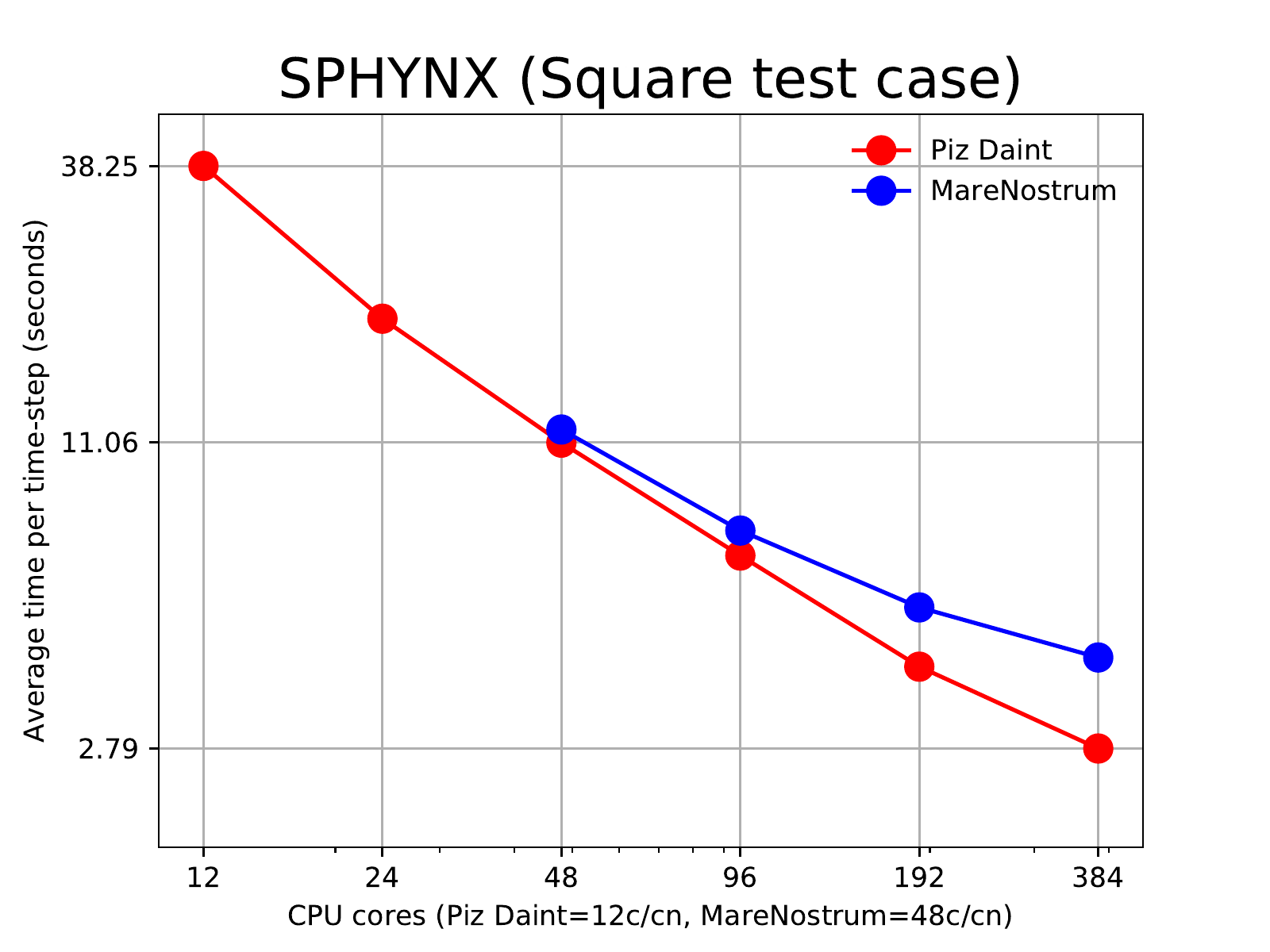}
  \label{fig:sphynx_square}
}%
\subfloat[]{
  \includegraphics[width=0.5\textwidth]{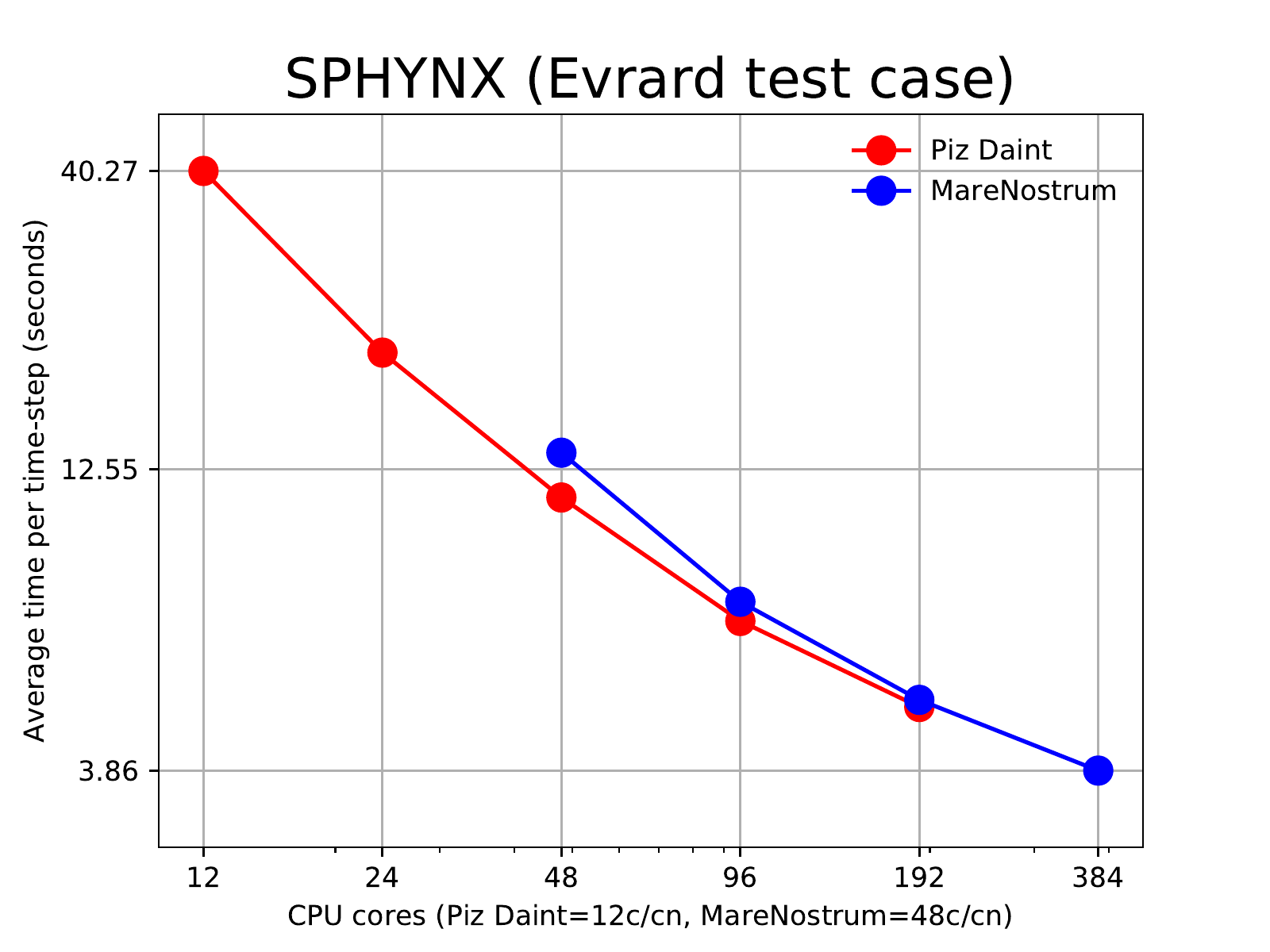}
  \label{fig:sphynx_evrard}
}%
\caption{Strong scalability results with SPHYNX.}
\label{fig:sphynx_strong}
\vspace{-1.2\baselineskip}
\end{figure*}

\begin{figure*}[h!]%
\centering
\subfloat[]{
  \includegraphics[width=0.35\textwidth]{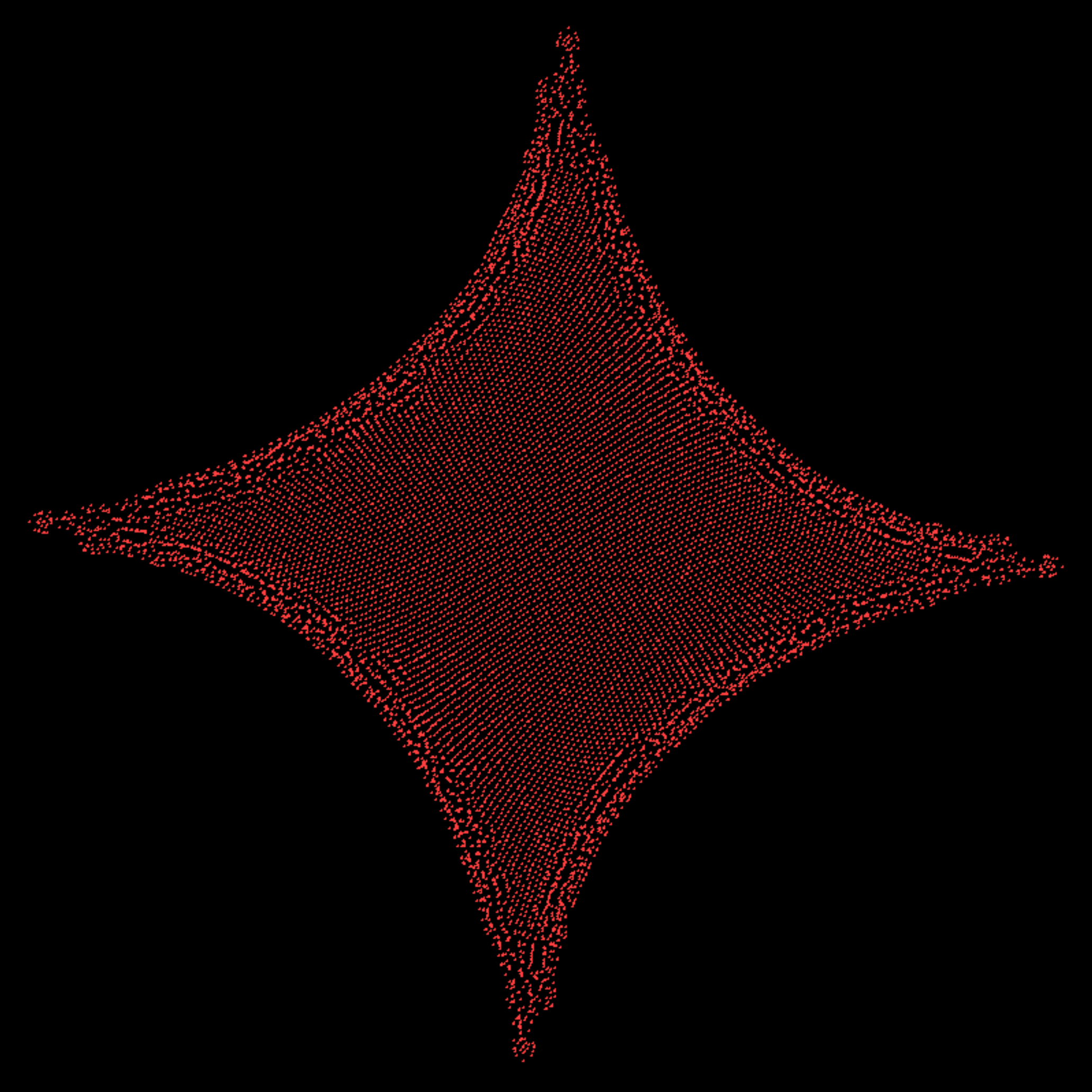}
  \label{tab:changa_pop}
}\hspace{0.5cm}%
\subfloat[]{
  \includegraphics[width=0.5\textwidth]{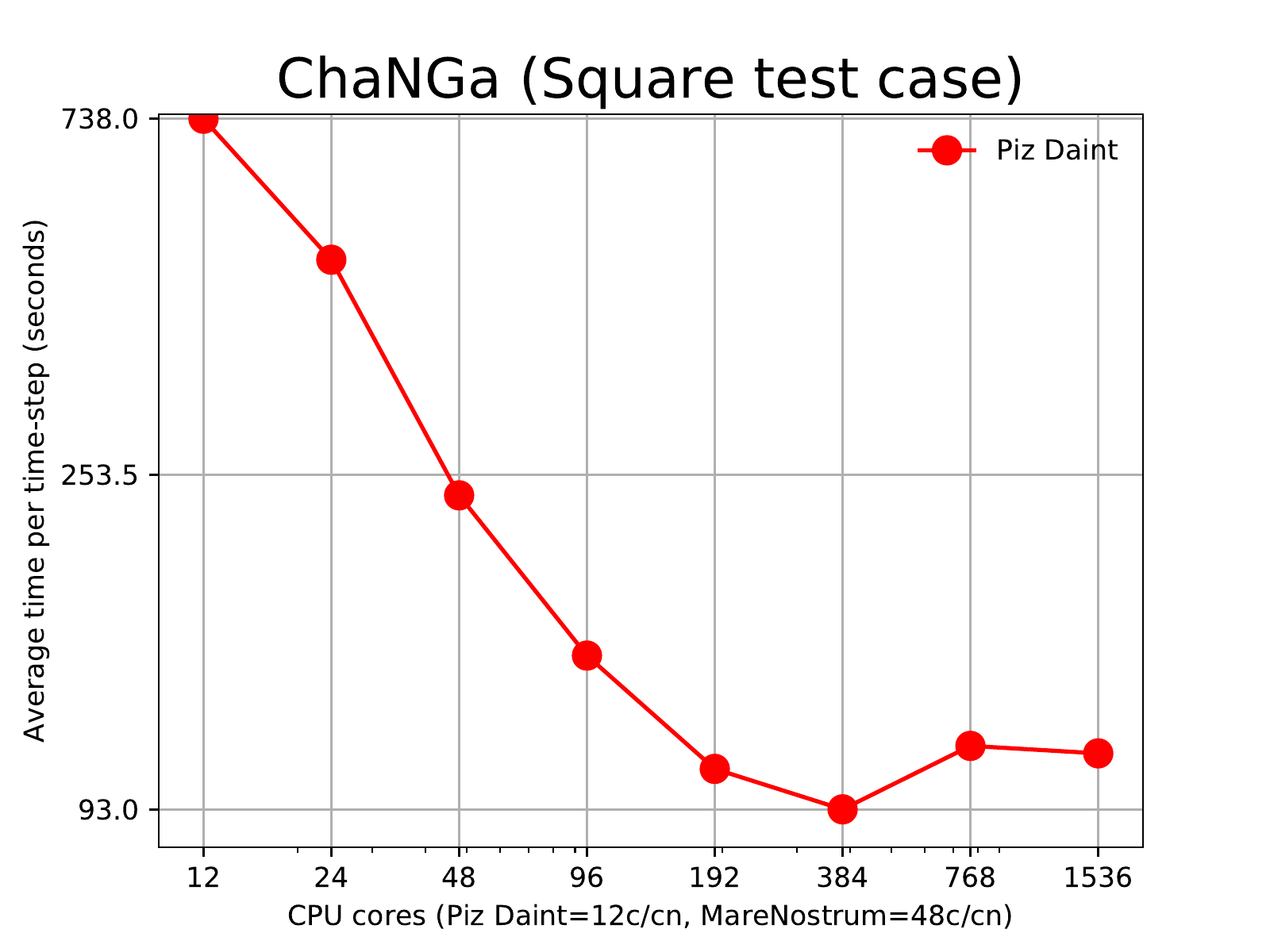}
  \label{fig:changa_square}
}%
\subfloat[]{
  \includegraphics[width=0.5\textwidth]{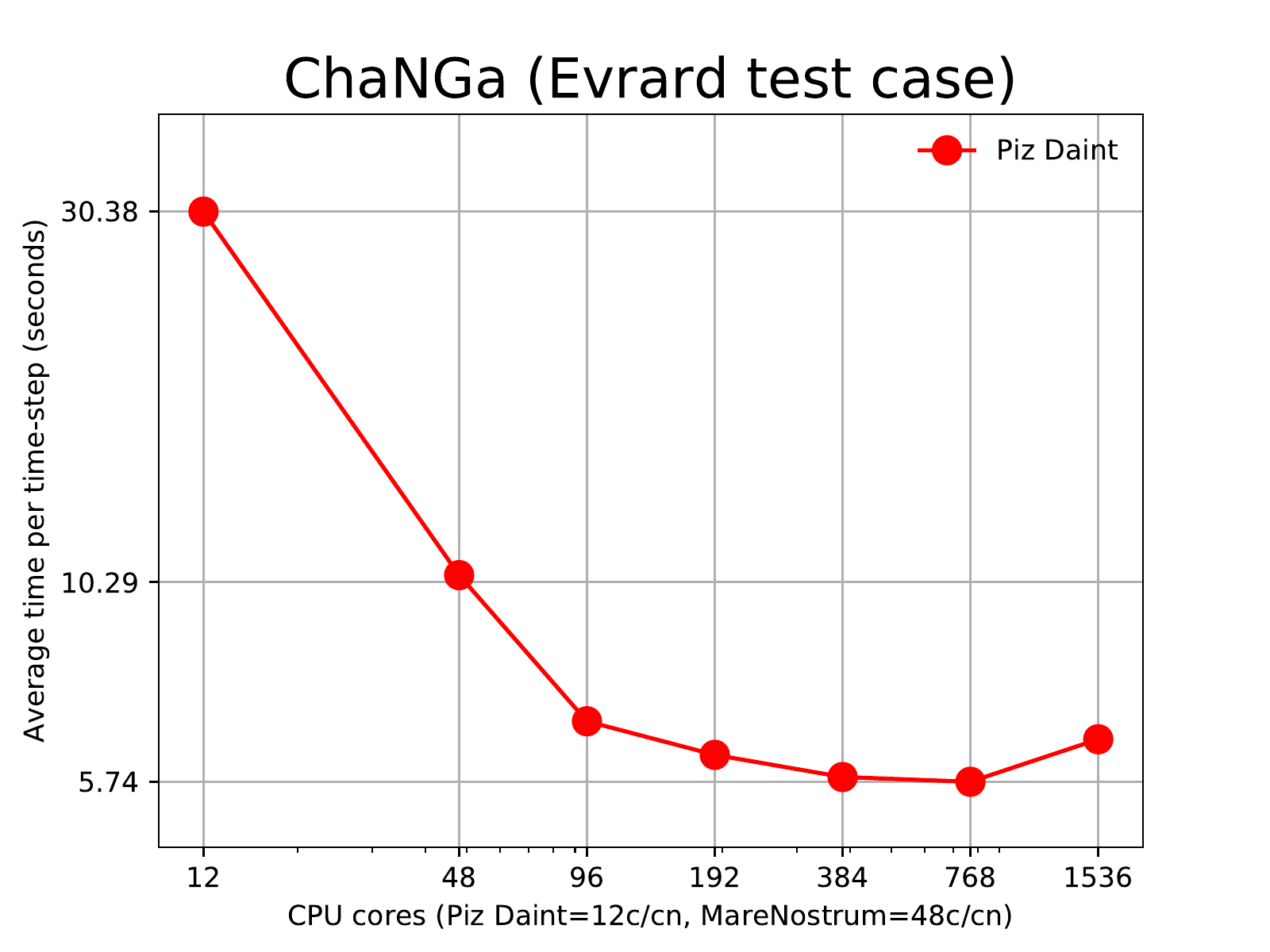}
  \label{fig:changa_evrard}
}%

\caption{Strong scalability results with ChaNGa.}
\label{fig:changa_strong}
\vspace{-0.5\baselineskip}
\end{figure*}

\begin{figure}%
\centering
  \includegraphics[width=0.61\columnwidth]{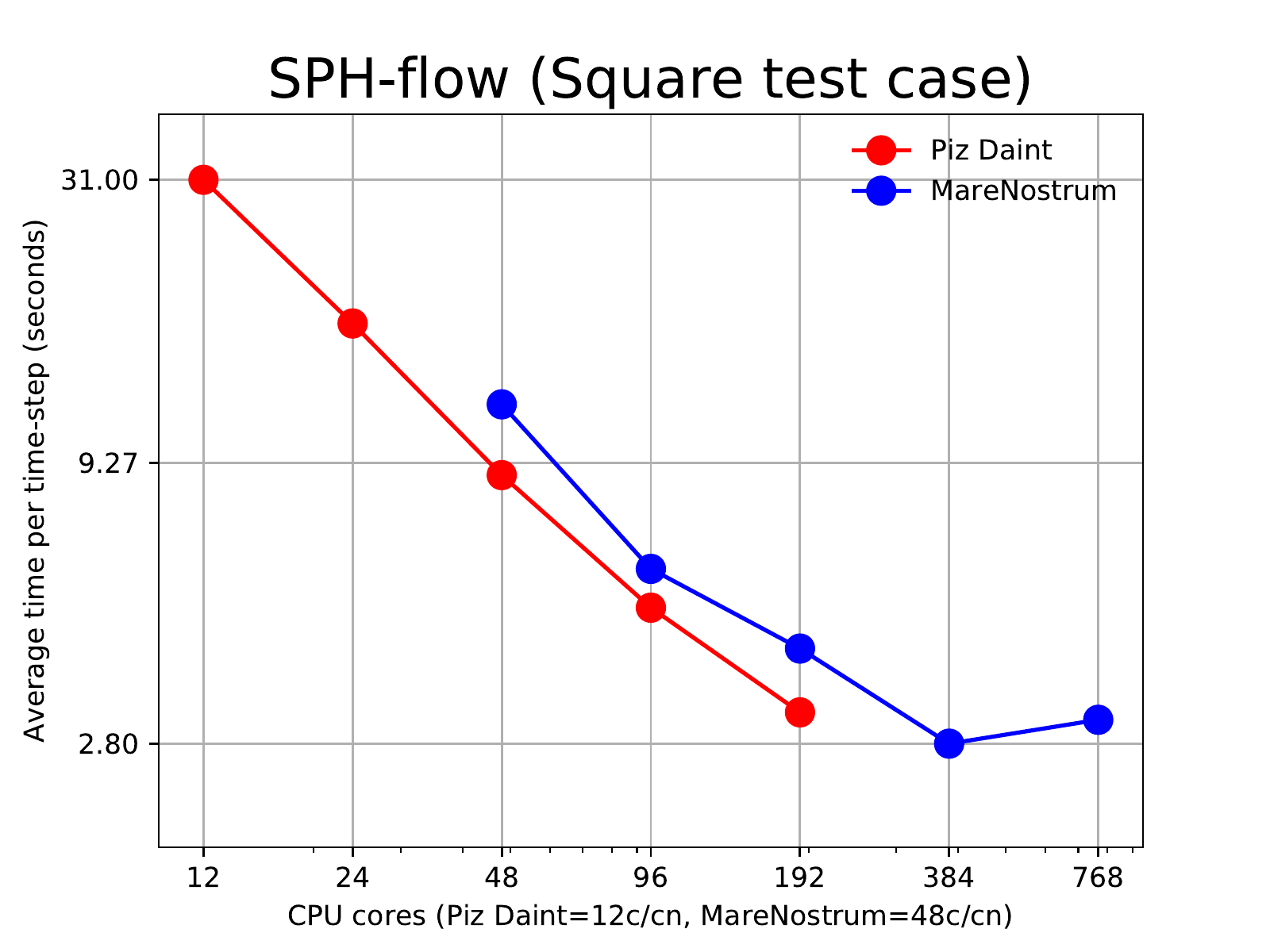}
  \label{fig:sphflow_square}
\caption{Strong scalability results with SPH-flow.} 
\label{fig:sphflow_strong}
\vspace{-0.5\baselineskip}
\end{figure}

Figures~\ref{fig:sphynx_strong},~\ref{fig:changa_strong} and \ref{fig:sphflow_strong} show the timings for SPHYNX, ChaNGa and SPH-flow. 
The red and blue lines show the average execution per time-step (in seconds) on Piz Daint and MareNostrum, respectively, up to 1,536 cores.
For the selected test cases (Evrard collapse and rotating square patch with $10^6$ particles), the applications exhibit good strong scaling up to 16 compute nodes on both systems. 
Scaling stalls when there are not enough particles/core (typically $10^4$) to keep the compute nodes busy.
However, realistic scientific simulations deal with both, more calculations per particle (i.e. detailed physics) and larger numbers of particles, where strong scaling is expected up to thousands of cores. 
We also stress here that the results with $10^6$ particles should not be generalized. 
This relatively low count of particles/core was chosen knowing that the codes will rapidly encounter a limit in their strong scaling due to reduced particles count per node with increasing node count. 
This easily exposed scalability limits of the codes and targets of interest for improvement. 
A factor that has not yet been explored is the weak scaling of these codes, which is usually the regime in which they operate in production runs. 
This is part of ongoing analysis work.

To help identify the root cause of the slowdown, we used the Extrae\footnote{\url{http://www.bsc.es/computer-sciences/performance-tools}} performance analysis tool. 
The tool provides a set of meaningful performance metrics to understand and guide the optimization of parallel codes. 
A way to compare the performance between the codes as well as its impact with variable core count is to examine a set of predefined performance metrics. 
Thereupon, efficiencies can be calculated from these metrics to identify which characteristics of the code contribute to performance inefficiencies.

\textit{Load Balance} is computed as the ratio between average useful computation time (across all processes) and maximum useful computation time (also across all processes). 

While the communication efficiency and computation scalability are close to ideal, the measured global efficiency steadily decreases from 48 cores to 192 cores. 
Most of the efficiency loss comes from an increased load imbalance, presumably caused by high idle times of certain worker tasks/threads. 

\subsubsection{Understanding implementation limits}

The experiments show that the selected parent implementations already suffer from load imbalance.  
Acknowledging that imbalance will become an even more important problem on exascale (-ready) machines, the SPH-EXA mini-app is being designed to take it into account, already in its early development phases, and addressing it via state-of-the-art, scalable dynamic load balancing mechanisms within a single compute node and across massive numbers of nodes. 

Additionally, fault-tolerance is currently being addressed via the combination of selective replication, algorithm-based fault-tolerance (ABFT) techniques, and optimal checkpointing, to sustain its scalable execution.

\subsubsection{Understanding SPHYNX execution}

Figure~\ref{fig:sphynx_strong_lb} shows the load balance of SPHYNX for a single time-step as measured for the Evrard collapse test with Extrae with 192 cores on Piz Daint. 
The different colors represent different main execution states: 
computing phases (blue), MPI collective communication (orange), thread synchronization (red), thread fork/join (yellow), and idle threads (black).
Each letter can be related to the different phases of Algorithm~\ref{fig.algo.sph}. 
Phase A is the building of the octree. 
Phases B, C, and D concern the finding of neighbors. 
Phases E to H are the SPH-related calculations (density, momentum, and energy, among other needed quantities). 
Phase I is the calculation of self-gravity. 
Finally, phase J, is the computation of the new time-step and the update of particle positions.

A highly scalable code will need not contain any of the black parallel regions (idle threads), and achieve that all blue regions are completed at the same time (load balancing). 
It is clear from Fig.~\ref{fig:sphynx_strong_lb} that the version 1.3.1 of SPHYNX examined therein had room for improvement. 
This study highlighted the importance of parallelizing the tree building (phase A), to address the idle regions (B, D, and J have been parallelized or re-written to be eliminated), and to include load balancing at intra- and inter-node level. 
The analysis and changes resulted in a more scalable SPHYNX version, which is currently under development.

{The results presented above were obtained within the framework of the present project. The performance behaviour was also confirmed by an independent audit of the performance data provided by us to the Performance Optimisation and Productivity (POP CoE)\footnote{\url{https://pop-coe.eu}} team. 
A similar performance behavior confirmation was conducted for SPH-flow.

\begin{figure}[!t]%
\centering
  \includegraphics[width=\columnwidth]{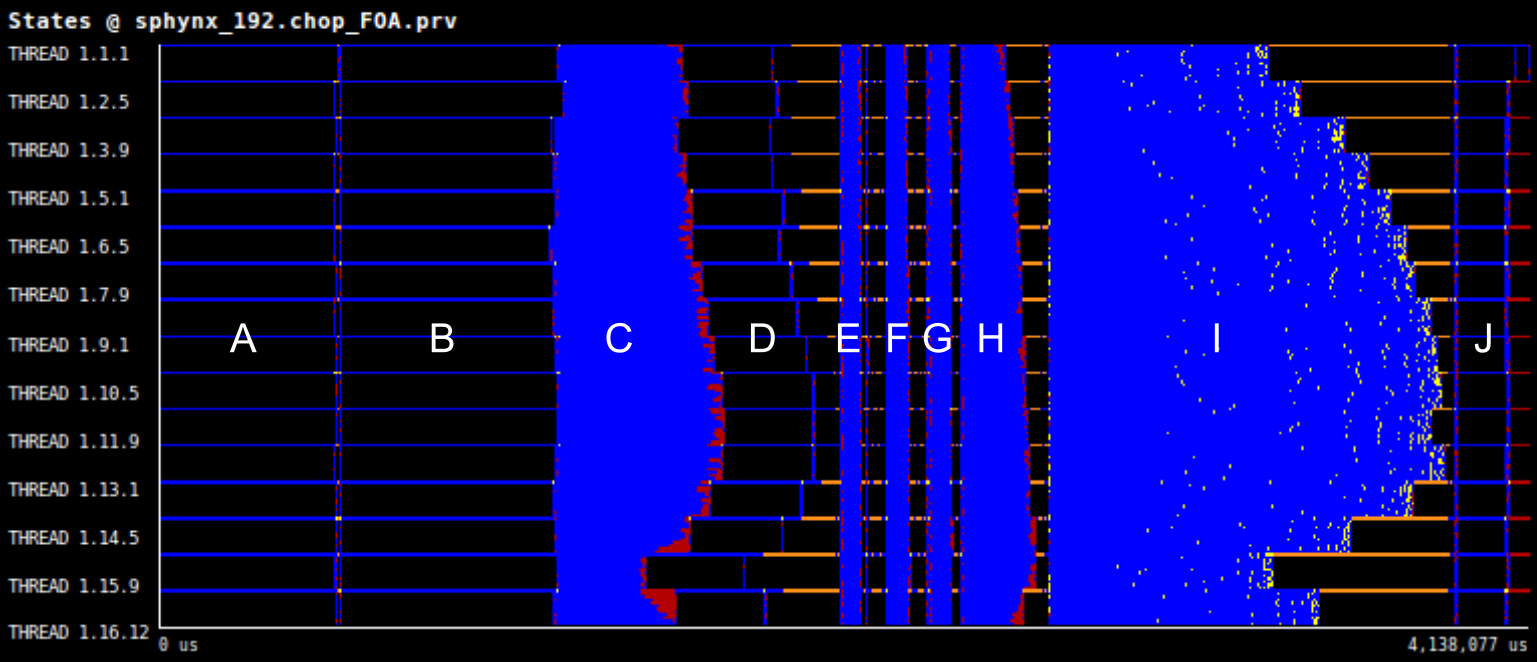}
\caption{Extrae visualization of SPHYNX v1.3.1 execution}
\label{fig:sphynx_strong_lb}
\vspace{-0.8\baselineskip}
\end{figure}

\section{Conclusion and Future Work}
\label{sec:conclusion}

An extensive study of three SPH implementations was performed in this work.
First, to gain insights in designing the SPH-EXA mini-app, we extrapolated the common basic SPH features of the three codes. 
Based on interdisciplinary co-design, we identified the differences and similarities between the codes, both in terms of computer science-related features and physical models implemented therein.
This results in a list of selected features, algorithms, and physics modules that need to be incorporated in the SPH-EXA mini-app.

Then, to expose any limitations of the codes that may represent a major performance degradation today or in the future Exascale era, we compared the codes through two common test cases that serve both, as a test now and as a validation and acceptance proof for the upcoming mini-app, in the form of reproducible experiments.
The result of this work is a deeper understanding of the three parent SPH codes, with a direct feedback to their developers, that already benefit from this work in terms of unveiling parallelization problems and improving overall scalability.

The expected outcome of the broader SPH-EXA project~\cite{SPH-EXA} will be in the form of an open-source SPH-EXA mini-app, that will enable highly parallelized, scalable, and fault-tolerant production SPH codes in different scientific domains.

\section*{Acknowledgment}
The authors thank Ioana Banicescu, Domingo García-Senz, and Thomas Quinn for fruitful discussions, valuable comments, and continuous support.

This work is supported in part by the Swiss Platform for Advanced Scientific Computing (PASC) project SPH-EXA~\cite{SPH-EXA} (2017- 2020).

The authors acknowledge the support of the Swiss National Supercomputing Centre (CSCS) via allocation project c16, Barcelona Supercomputing Center (BSC), and the Center for scientific computing at the University of Basel (sciCORE) where the calculations were performed.

Certain computation and simulation results presented here were obtained with SPH-flow, courtesy of Nextflow Software.
The support of the EU Horizon 2020 POP CoE project is also acknowledged.

\bibliographystyle{abbrv}
\bibliography{ref}

\end{document}